\pdfoutput=1 
\documentclass[12pt,a4paper]{article}

\usepackage{ifthen} 
\newboolean{pdflatex}
\setboolean{pdflatex}{true} 

\newboolean{articletitles}
\setboolean{articletitles}{true} 

\newboolean{uprightparticles}
\setboolean{uprightparticles}{false} 
\usepackage[normalem]{ulem}

\def\paperauthors{LHCb collaboration} 
\def\paperasciititle{Measurement of the Psi2s to Jpsi cross-section ratio as a function of centrality in PbPb collisions at sqrt(sNN) = 5.02 TeV} 
\def\papertitle{Measurement of the \psitwos to \jpsi cross-section ratio as a function of centrality in PbPb collisions at $\sqsnn = 5.02\tev$} 
\def\paperkeywords{{High Energy Physics}, {LHCb}} 
\def\papercopyright{\the\year\ CERN for the benefit of the LHCb collaboration} 
\def\paperlicence{CC BY 4.0 licence}
\def\paperlicenceurl{https://creativecommons.org/licenses/by/4.0/}

\newif\ifEnableSectionTOCLinks
\EnableSectionTOCLinksfalse 

\usepackage[top=1in, bottom=1.25in, left=1in, right=1in]{geometry}

\usepackage{microtype}
\usepackage{lineno}  
\usepackage{xspace} 
\usepackage{caption} 

\usepackage{graphicx}  
\usepackage{color}
\usepackage{colortbl}
\graphicspath{{./figs/}} 

\usepackage{amsmath} 
\usepackage{amssymb}
\usepackage{amsfonts}
\usepackage{upgreek} 

\newcommand*\patchAmsMathEnvironmentForLineno[1]{%
\expandafter\let\csname old#1\expandafter\endcsname\csname #1\endcsname
\expandafter\let\csname oldend#1\expandafter\endcsname\csname
end#1\endcsname
 \renewenvironment{#1}%
   {\linenomath\csname old#1\endcsname}%
   {\csname oldend#1\endcsname\endlinenomath}%
}
\newcommand*\patchBothAmsMathEnvironmentsForLineno[1]{%
  \patchAmsMathEnvironmentForLineno{#1}%
  \patchAmsMathEnvironmentForLineno{#1*}%
}
\AtBeginDocument{%
\patchBothAmsMathEnvironmentsForLineno{equation}%
\patchBothAmsMathEnvironmentsForLineno{align}%
\patchBothAmsMathEnvironmentsForLineno{flalign}%
\patchBothAmsMathEnvironmentsForLineno{alignat}%
\patchBothAmsMathEnvironmentsForLineno{gather}%
\patchBothAmsMathEnvironmentsForLineno{multline}%
\patchBothAmsMathEnvironmentsForLineno{eqnarray}%
}

\usepackage{hyperref}
\usepackage{hyperxmp}
\usepackage{hyperref}[pdftex,
            pdfauthor={\paperauthors},
            pdftitle={\paperasciititle},
            pdfkeywords={\paperkeywords},
            pdfcopyright={Copyright (C) \papercopyright},
            pdflicenseurl={\paperlicenceurl}]

\usepackage[colorinlistoftodos,textsize=scriptsize]{todonotes}

\usepackage[bottom,flushmargin,hang,multiple]{footmisc}

\usepackage[all]{hypcap} 

\usepackage{xspace} 
\usepackage{upgreek}

\def\lhcb   {\mbox{LHCb}\xspace}

\def\MagUp {\mbox{\em Mag\kern -0.05em Up}\xspace}

\ifthenelse{\boolean{uprightparticles}}%
{

 \def\Pmu         {\ensuremath{\upmu}\xspace}

 \def\Ppsi        {\ensuremath{\uppsi}\xspace}

 \def\PDelta      {\ensuremath{\Delta}\xspace}                 
 \def\PXi         {\ensuremath{\Xi}\xspace}                 
 \def\PLambda     {\ensuremath{\Lambda}\xspace}                 
 \def\PSigma      {\ensuremath{\Sigma}\xspace}                 
 \def\POmega      {\ensuremath{\Omega}\xspace}                 
 \def\PUpsilon    {\ensuremath{\Upsilon}\xspace}
 \let\oldPi\Pi
 \def\PPi         {\ensuremath{\oldPi}\xspace}

 \def\PB      {\ensuremath{\mathrm{B}}\xspace}                 
 \def\PD      {\ensuremath{\mathrm{D}}\xspace}                 
 \def\PJ      {\ensuremath{\mathrm{J}}\xspace}                 
 \def\PK      {\ensuremath{\mathrm{K}}\xspace}                 
 \def\Pb      {\ensuremath{\mathrm{b}}\xspace}                 
 \def\Pc      {\ensuremath{\mathrm{c}}\xspace}

 \def\Ps      {\ensuremath{\mathrm{s}}\xspace}

 \def\thebaroffset{0.0em}
}
{

 \def\Pmu         {\ensuremath{\mu}\xspace}

 \def\Ppsi        {\ensuremath{\psi}\xspace}                 
                  
 \mathchardef\PDelta="7101
 \mathchardef\PXi="7104
 \mathchardef\PLambda="7103
 \mathchardef\PSigma="7106
 \mathchardef\POmega="710A
 \mathchardef\PUpsilon="7107
 \mathchardef\PPi="7105
 \def\PB      {\ensuremath{B}\xspace}                 
 \def\PD      {\ensuremath{D}\xspace}                 
 \def\PJ      {\ensuremath{J}\xspace}                 
 \def\PK      {\ensuremath{K}\xspace}                 
 \def\Pb      {\ensuremath{b}\xspace}                 
 \def\Pc      {\ensuremath{c}\xspace}

 \def\Ps      {\ensuremath{s}\xspace}

 \def\thebaroffset{0.18em}
}
\newcommand{\offsetoverline}[2][\thebaroffset]{\kern #1\overline{\kern -#1 #2}}%

\makeatletter
\ifcase \@ptsize \relax
  \newcommand{\miniscule}{\@setfontsize\miniscule{4}{5}}
\or
  \newcommand{\miniscule}{\@setfontsize\miniscule{5}{6}}
\or
  \newcommand{\miniscule}{\@setfontsize\miniscule{5}{6}}
\fi
\makeatother

\DeclareRobustCommand{\optbar}[1]{\shortstack{{\miniscule (\rule[.5ex]{1.25em}{.18mm})}
  \\ [-.7ex] $#1$}}

\def\mup        {{\ensuremath{\Pmu^+}}\xspace}
\def\mun        {{\ensuremath{\Pmu^-}}\xspace} 

\def\squark    {{\ensuremath{\Ps}}\xspace}

\def\cquark    {{\ensuremath{\Pc}}\xspace}
\def\cquarkbar {{\ensuremath{\overline \cquark}}\xspace}
\def\ccbar     {{\ensuremath{\cquark\cquarkbar}}\xspace}
\def\bquark    {{\ensuremath{\Pb}}\xspace}

\def\KorKbar {\kern \thebaroffset\optbar{\kern -\thebaroffset \PK}{}\xspace}

\def\D       {{\ensuremath{\PD}}\xspace}

\def\DorDbar {\kern \thebaroffset\optbar{\kern -\thebaroffset \PD}\xspace}

\def\Dp      {{\ensuremath{\D^+}}\xspace}
\def\Dm      {{\ensuremath{\D^-}}\xspace}

\def\DpDm    {\ensuremath{\Dp {\kern -0.16em \Dm}}\xspace}

\def\B       {{\ensuremath{\PB}}\xspace}

\def\BorBbar {\kern \thebaroffset\optbar{\kern -\thebaroffset \PB}\xspace}

\def\Bd      {{\ensuremath{\B^0}}\xspace}

\def\BdorBdbar {\kern \thebaroffset\optbar{\kern -\thebaroffset \Bd}\xspace}

\def\Bs      {{\ensuremath{\B^0_\squark}}\xspace}

\def\BsorBsbar {\kern \thebaroffset\optbar{\kern -\thebaroffset \Bs}\xspace}

\def\jpsi     {{\ensuremath{{\PJ\mskip -3mu/\mskip -2mu\Ppsi}}}\xspace}
\def\psitwos  {{\ensuremath{\Ppsi{(2S)}}}\xspace}

\def\Y#1S{\ensuremath{\PUpsilon{(#1S)}}\xspace}

\def\LorLbar     {\kern \thebaroffset\optbar{\kern -\thebaroffset \PLambda}\xspace}

\def\BF         {{\ensuremath{\mathcal{B}}}\xspace}

\def\to                 {\ensuremath{\rightarrow}\xspace}

\def\AT#1     {\ensuremath{A_{\mathrm{T}}^{#1}}\xspace}           

\def\C#1      {\ensuremath{\mathcal{C}_{#1}}\xspace}                       
\def\Cp#1     {\ensuremath{\mathcal{C}_{#1}^{'}}\xspace}                    
\def\Ceff#1   {\ensuremath{\mathcal{C}_{#1}^{\mathrm{(eff)}}}\xspace}        
\def\Cpeff#1  {\ensuremath{\mathcal{C}_{#1}^{'\mathrm{(eff)}}}\xspace}       
\def\Ope#1    {\ensuremath{\mathcal{O}_{#1}}\xspace}                       
\def\Opep#1   {\ensuremath{\mathcal{O}_{#1}^{'}}\xspace}                    

\newcommand{\nospaceunit}[1]{\ensuremath{\text{#1}}}       
\newcommand{\aunit}[1]{\ensuremath{\text{\,#1}}}       

\newcommand{\tev}{\aunit{Te\kern -0.1em V}\xspace}
\newcommand{\gev}{\aunit{Ge\kern -0.1em V}\xspace}
\newcommand{\mev}{\aunit{Me\kern -0.1em V}\xspace}
\newcommand{\kev}{\aunit{ke\kern -0.1em V}\xspace}
\newcommand{\ev}{\aunit{e\kern -0.1em V}\xspace}
 
\newcommand{\mevc}{\ensuremath{\aunit{Me\kern -0.1em V\!/}c}\xspace}
\newcommand{\gevc}{\ensuremath{\aunit{Ge\kern -0.1em V\!/}c}\xspace}
\newcommand{\mevcc}{\ensuremath{\aunit{Me\kern -0.1em V\!/}c^2}\xspace}
\newcommand{\gevcc}{\ensuremath{\aunit{Ge\kern -0.1em V\!/}c^2}\xspace}

\def\mum  {\ensuremath{\,\upmu\nospaceunit{m}}\xspace}

\newcommand{\chisq}{\ensuremath{\chi^2}\xspace}

\newcommand{\chisqip}{\ensuremath{\chi^2_{\text{IP}}}\xspace}

\def\gsim{{~\raise.15em\hbox{$>$}\kern-.85em
          \lower.35em\hbox{$\sim$}~}\xspace}
\def\lsim{{~\raise.15em\hbox{$<$}\kern-.85em
          \lower.35em\hbox{$\sim$}~}\xspace}

\def\sqsnn {\ensuremath{\protect\sqrt{s_{\scriptscriptstyle\text{NN}}}}\xspace}
\def\pt         {\ensuremath{p_{\mathrm{T}}}\xspace}

\def\evtgen     {\mbox{\textsc{EvtGen}}\xspace}

\def\pythia     {\mbox{\textsc{Pythia}}\xspace}

\def\tell1  {TELL1\xspace}
\def\ukl1   {UKL1\xspace}

\newcommand{\lhcborcid}[1]{\href{https://orcid.org/#1}{\hspace*{0.1em}\raisebox{-0.45ex}{\includegraphics[width=1em]{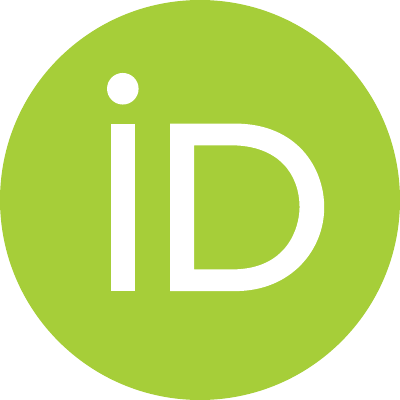}}}}

\usepackage{cite} 
\usepackage{mciteplus}

\usepackage{longtable} 
\begin{document}

\renewcommand{\thefootnote}{\fnsymbol{footnote}}
\setcounter{footnote}{1}


\begin{titlepage}
\pagenumbering{roman}

\vspace*{-1.5cm}
\centerline{\large EUROPEAN ORGANIZATION FOR NUCLEAR RESEARCH (CERN)}
\vspace*{1.5cm}
\noindent
\begin{tabular*}{\linewidth}{lc@{\extracolsep{\fill}}r@{\extracolsep{0pt}}}
\ifthenelse{\boolean{pdflatex}}
{\vspace*{-1.5cm}\mbox{\!\!\!\includegraphics[width=.14\textwidth]{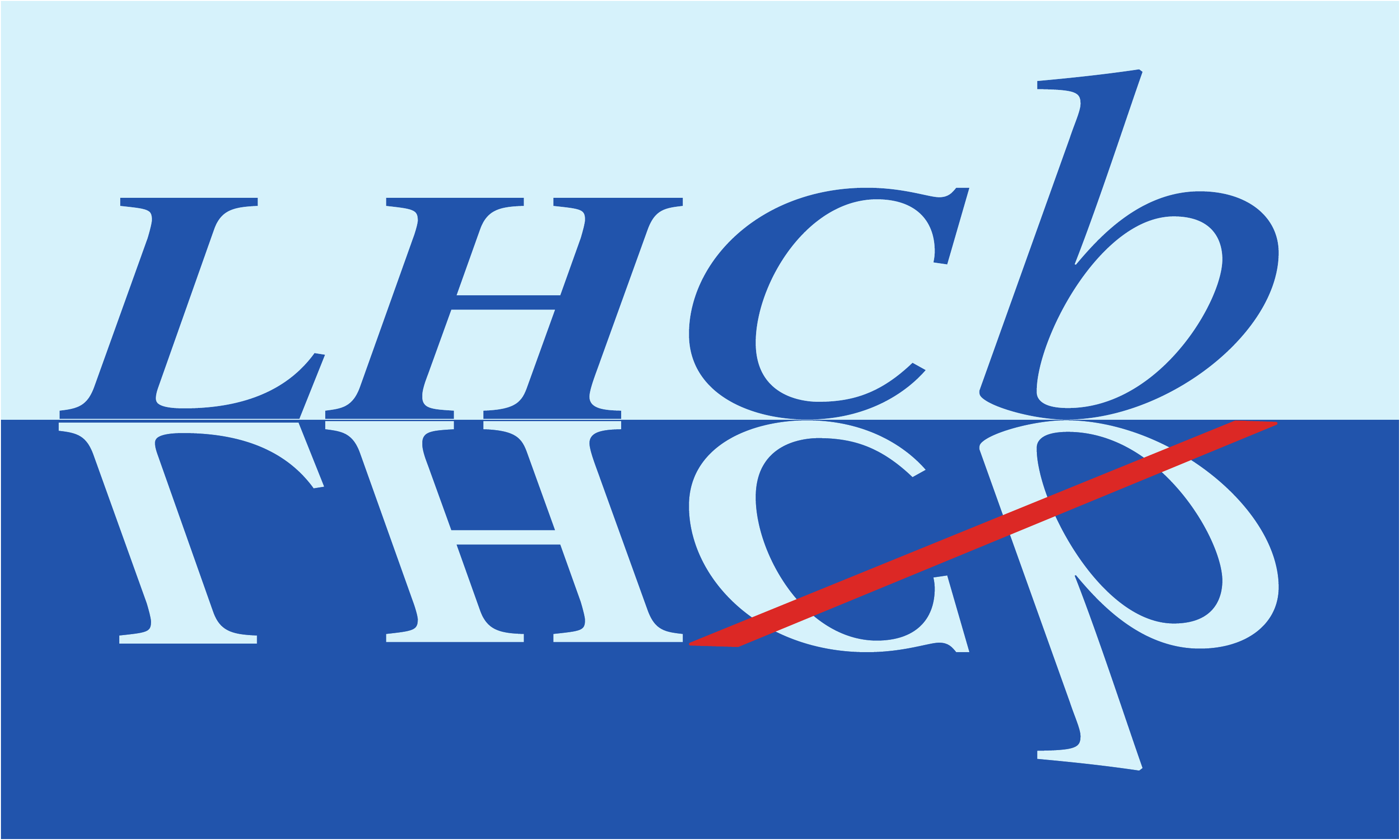}} & &}%
{\vspace*{-1.2cm}\mbox{\!\!\!\includegraphics[width=.12\textwidth]{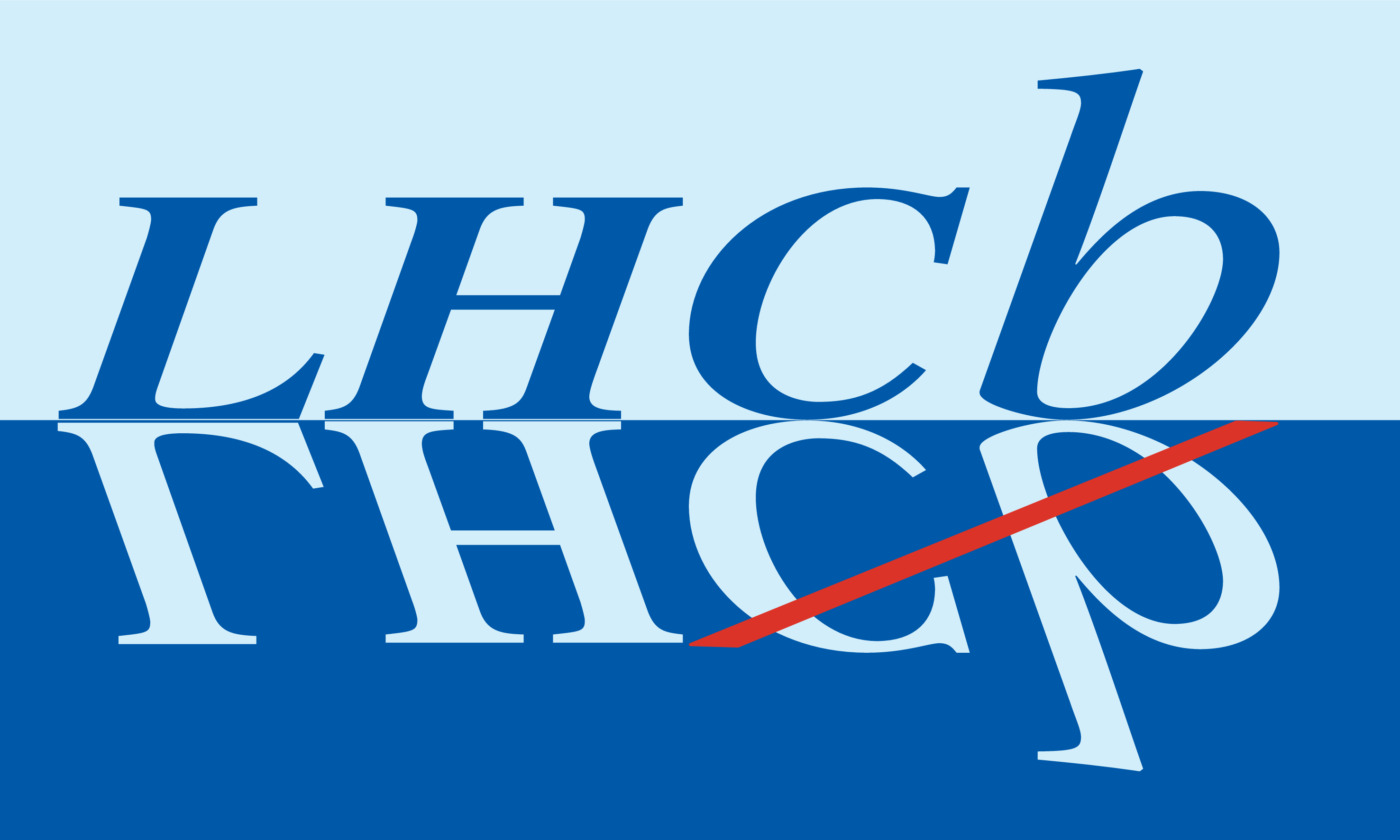}} & &}%
\\
 & & CERN-EP-2024-272 \\  
 & & LHCb-PAPER-2024-041 \\  
 & & November 7, 2024 \\ 
 & & \\
\end{tabular*}

\vspace*{4.0cm}

{\normalfont\bfseries\boldmath\huge
\begin{center}
  \papertitle 
\end{center}
}

\vspace*{2.0cm}

\begin{center}
\paperauthors\footnote{Authors are listed at the end of this paper.}
\end{center}

\vspace{\fill}

\begin{abstract}
  \noindent
   The ratio of prompt production cross-sections of $\ensuremath{\Ppsi{(2S)}}$ and $\ensuremath{{\PJ\mskip -3mu/\mskip -2mu\Ppsi}}$ mesons in their dimuon final state is measured as a function of centrality, using data collected by the LHCb detector in PbPb collisions at $\ensuremath{\protect\sqrt{s_{\scriptscriptstyle\text{NN}}}}$ = 5.02$\aunit{Te\kern -0.1em V}$, for the first time in the forward rapidity region.
   The measured ratio shows no dependence on the collision centrality, and is compared to the latest theory predictions and to the recent measurements in literature.
\end{abstract}

\vspace*{2.0cm}

\begin{center}
  Published in JHEP07(2025)235 
\end{center}

\vspace{\fill}

{\footnotesize 
\centerline{\copyright~\papercopyright. \href{\paperlicenceurl}{\paperlicence}.}}
\vspace*{2mm}

\end{titlepage}


\newpage
\setcounter{page}{2}
\mbox{~}
%
%
%
%


\renewcommand{\thefootnote}{\arabic{footnote}}
\setcounter{footnote}{0}


\cleardoublepage


\pagestyle{plain} 
\setcounter{page}{1}
\pagenumbering{arabic}


\section{Introduction}
\label{sec:Introduction}
The production of charmonia, in particular of the \jpsi and \psitwos states, has long been considered a key probe for understanding the properties of the quark-gluon plasma (QGP).
The idea that quarkonium suppression is an indicator of QGP formation was introduced by T.~Matsui and H.~Satz, who proposed that in deconfined nuclear matter, \ccbar pairs become unbound under the effect of colour screening~\cite{Matsui:1986dk}, 
with different yields for different states according to their binding energy.
Recently, the regeneration mechanism, within the QGP or at the phase boundary, has been considered an important ingredient for describing the \jpsi yields at the LHC~\cite{ALICE:2016flj} and nominal RHIC energies~\cite{PhysRevLett.97.232301}, with \psitwos production relative to that of \jpsi representing one possible discriminator between the two different regeneration scenarios~\cite{Braun-Munzinger:2000csl, Thews:2000rj}. 
Early results from the NA50 collaboration at the SPS indicated \jpsi suppression in
PbPb collisions~\cite{NA50:1997rtj, NA50:2004sgj}. These results were extensively debated but were not a conclusive signature of QGP formation as other competing mechanisms, such as final-state interactions with comoving particles, can contribute to quarkonium suppression~\cite{Ferreiro:2012rq}.  
Further studies in small systems by the E866~\cite{Leitch:2000jpsi}, PHENIX~\cite{PHENIX:2022nrm}, ALICE~\cite{ALICE:2014cgk}, CMS~\cite{CMS:2018gbb} and LHCb~\cite{LHCb-PAPER-2015-058,LHCb-PAPER-2017-014} experiments showed that the charmonium suppression is observed even in collision systems like proton-nucleus ($p$A).
In particular, the suppression of \psitwos production relative to \jpsi in $p$A collisions suggests that comovers mechanisms are important in understanding quarkonium behaviour in high-energy collisions. 
Theoretical models accounting for these effects have been developed to provide a more comprehensive interpretation of the data~\cite{Ferreiro:2014bia}.
In addition, results from the PHENIX~\cite{PHENIX:2006gsi} and STAR~\cite{STAR:2019fge}
experiments demonstrated significant
suppression of \jpsi yields in AuAu collisions, providing insight into the competing effects of QGP formation and CNM contributions. These findings were then complemented by the ALICE~\cite{ALICE:2016flj} and CMS~\cite{CMS:2012bms} experiments which studied charmonium production in hadronic PbPb collisions. Furthermore, the ALICE~\cite{ALICE:2021gpt}, CMS~\cite{CMS:2023snh}, and LHCb~\cite{LHCb-PAPER-2022-012} collaborations have investigated charmonium photoproduction in ultra-peripheral collisions (UPCs), offering a complementary perspective by probing the gluon distributions in nuclei at low Bjorken-x.
Together, these studies have advanced the understanding of the interplay between cold and hot nuclear matter effects,
motivating new investigations of the properties of the medium through the relative suppression of quarkonium states.

In this paper, the relative prompt production of \psitwos to \jpsi states, both decaying to two muons, is studied as a function of centrality using the 2018 PbPb collision data collected at a nucleon-nucleon centre-of-mass energy of $\sqsnn=5.02\tev$ by the \lhcb experiment, corresponding to an integrated luminosity of around $230 \, \upmu\rm{b}^{-1}$~\cite{LHCb-PAPER-2014-047}. 
The ratio of cross-sections ($\sigma$) of the two charmonium states multiplied by their branching fractions ($\mathcal{B}$) in the dimuon channel is measured as
\begin{equation}
\label{eq:ratio}
\dfrac{\BF(\psitwos\to\mup\mun)}{\BF(J/\psi\to\mup \mun)}\cdot\dfrac{\sigma(\psitwos)}{\sigma(J/\psi)} = \dfrac{N(\psitwos)}{N(J/\psi)}\cdot 
\dfrac{\varepsilon_\mathrm{tot}(J/\psi)}
{\varepsilon_\mathrm{tot}(\psitwos)},
\end{equation}
where $N$ and $\varepsilon_\mathrm{tot}$ are the signal yields and efficiencies, respectively.

\section{The LHCb detector and data selection}\label{sec:datasel}
The LHCb detector~\cite{LHCb-DP-2008-001,LHCb-DP-2014-002} is a single-arm forward spectrometer covering the pseudorapidity range 2 $< \eta <$ 5, designed for the study of particles containing \bquark\ or \cquark\ quarks. The detector includes a high-precision tracking system consisting of a silicon-strip vertex detector (VELO) surrounding the interaction region~\cite{Aaij:2014zzy}, a large-area silicon-strip detector located upstream of a dipole magnet with a bending power of about $4{\mathrm{\,T\,m}}$, and three stations of silicon-strip detectors and straw drift tubes~\cite{LHCbOuterTrackerGroup:2017hwu} placed downstream of the magnet. The tracking system provides a measurement of the momentum, $p$, of charged particles with a relative uncertainty that varies from 0.5\% at low momentum to 1.0\% at 200\gevc. The minimum distance of a track to a primary vertex (PV), the impact parameter, is measured with a resolution of $(15+29/\pt)\mum$, where \pt is the component of the momentum transverse to the beam, in \gevc. Different types of charged hadrons are distinguished using information from two ring-imaging Cherenkov detectors (RICH)~\cite{LHCbRICHGroup:2012mgd}. Photons, electrons and hadrons are identified by a calorimeter system consisting of scintillating-pad and preshower detectors, an electromagnetic calorimeter (ECAL) and a hadronic calorimeter. Muons are identified by a system composed of alternating layers of iron and multiwire proportional chambers~\cite{Alves:2012ey}.

The online event selection is performed by a trigger~\cite{LHCb-DP-2019-001}, which consists of a hardware stage, based on information from the calorimeter and muon systems, followed by a software stage, which applies a full event reconstruction. 
At the hardware level, events are required to have a muon with $\pt > 500 \mevc$. Events are selected at software level if the number of clusters in the VELO, $N_{\rm{c}}$, satisfies $6000 <$ $ N_{\rm{c}}$ $ < 10000$ or if $N_{\rm{c}}$ $ < 6000$ and two muons with $\pt > 700 \mevc$ are identified. 
Moreover, to suppress contamination from Pb-gas interactions,\footnote{Neon gas was injected in the beam pipe near the interaction point, using the LHCb fixed-target SMOG system~\cite{LHCb-PAPER-2014-047}, concurrently with PbPb collisions.} a minimum of 15 VELO tracks in the backward direction, $\eta < 0$, is required.
In order to select only the prompt contribution, each track must be consistent with originating from a PV, through a selection on the \chisqip, defined as the difference in the vertex-fit \chisq of a given PV reconstructed with and without the track under consideration. The remaining contamination from nonprompt events is found to be around 5\% for \jpsi and 8\% for \psitwos, and included in the systematic uncertainties on the ratio measurement. The \jpsi and \psitwos candidates are reconstructed in the dimuon final state, satisfying requirements on rapidity, $2.0 < y < 4.5$, and transverse momentum, $0.3 < \pt <  10 \gevc$, where the lower end of the \pt selection is chosen to reject photoproduced charmonium events~\cite{LHCb:2021hoq}.

Finally, the two muons are both required to have \pt greater than $900\mevc$ and to satisfy particle identification (PID) requirements based on information from the RICH, calorimeter and muon systems. 

\section{Yields in centrality intervals}
\label{sec:yield}
The charmonium yields are determined in intervals of the collision centrality. Centrality intervals are defined as percentiles of the total inelastic hadronic PbPb cross-section and are associated with the impact parameter of the colliding particles and the mean number of participating nucleons, $\langle N_{\rm{part}} \rangle$~\cite{LHCb:2021ysy}. A higher centrality percentile corresponds to more peripheral collisions with a larger impact parameter, while a lower centrality percentile corresponds to more central collisions with a smaller impact parameter.
To estimate these geometric quantities, the Glauber Monte Carlo (GMC) model adapted to LHCb conditions is used~\cite{LHCb:2021ysy,Loizides:2014vua}.
The impact parameter cannot be measured directly and is instead estimated using the total energy deposited in the ECAL, with which it scales linearly.
The method is based on a binned fit of the total ECAL energy in simulated minimum-bias PbPb interactions where the GMC model is used and the experimental conditions of the signal sample are reproduced. This fit establishes a per-event dependence of $\langle N_{\rm{part}} \rangle$ and the collision centrality in terms of the energy deposited in the ECAL.
The centrality intervals used are defined in Table~\ref{tab:mycen}.
The data used to determine the ratio are limited to centrality values down to about $60 \,\%$ due to the inability of the detector to cope with the increasing occupancy.

\begin{figure}[tb]
  \begin{center}
    \includegraphics[width=0.49\linewidth]{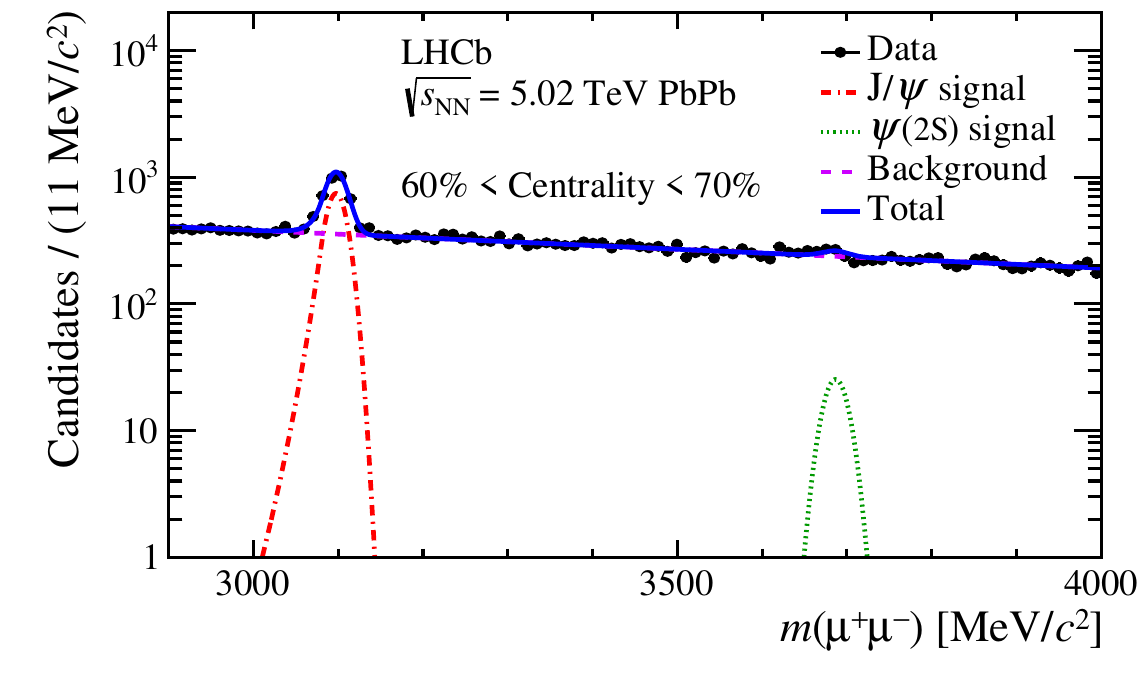}
    \includegraphics[width=0.49\linewidth]{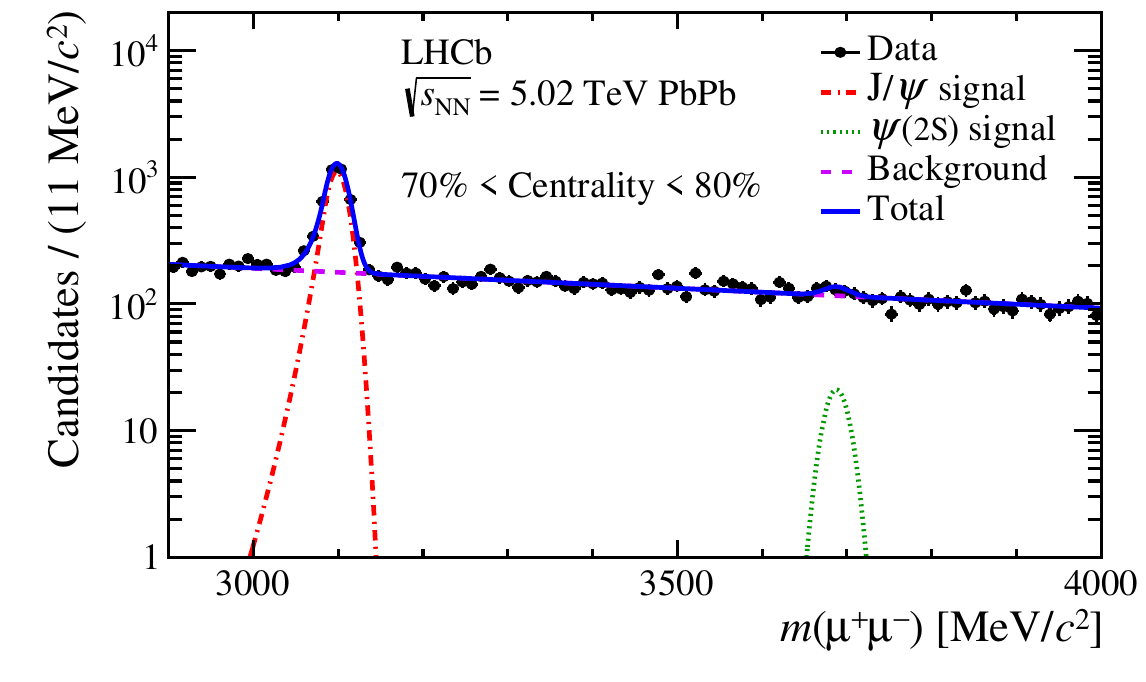}
    \includegraphics[width=0.49\linewidth]{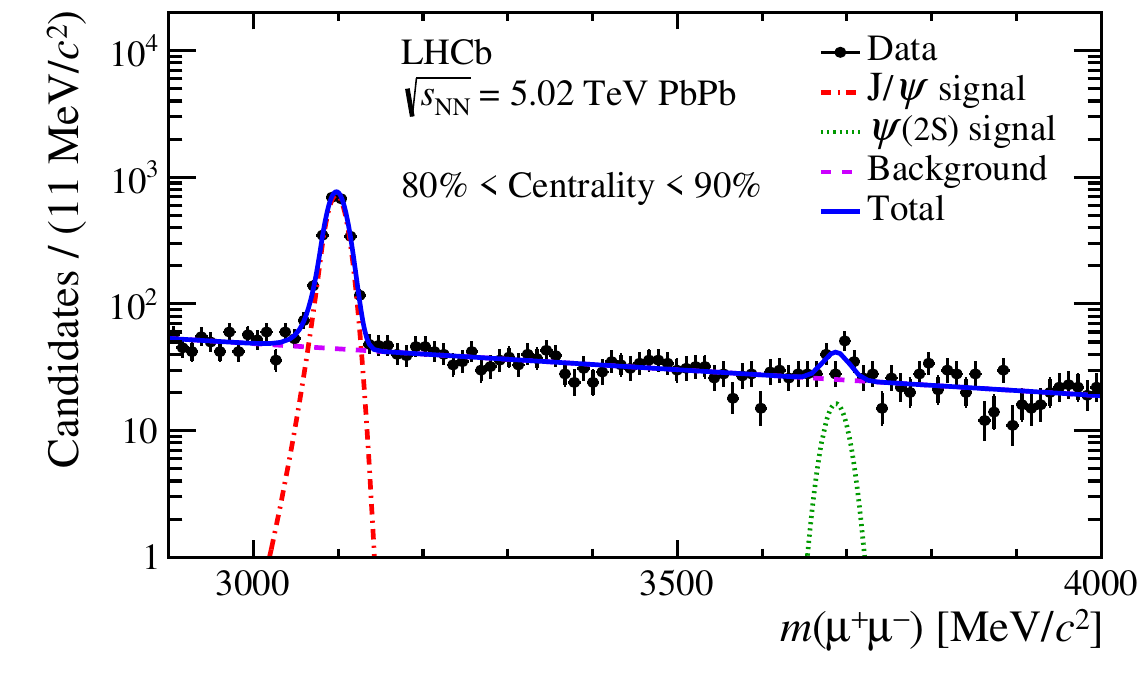}
    \includegraphics[width=0.49\linewidth]{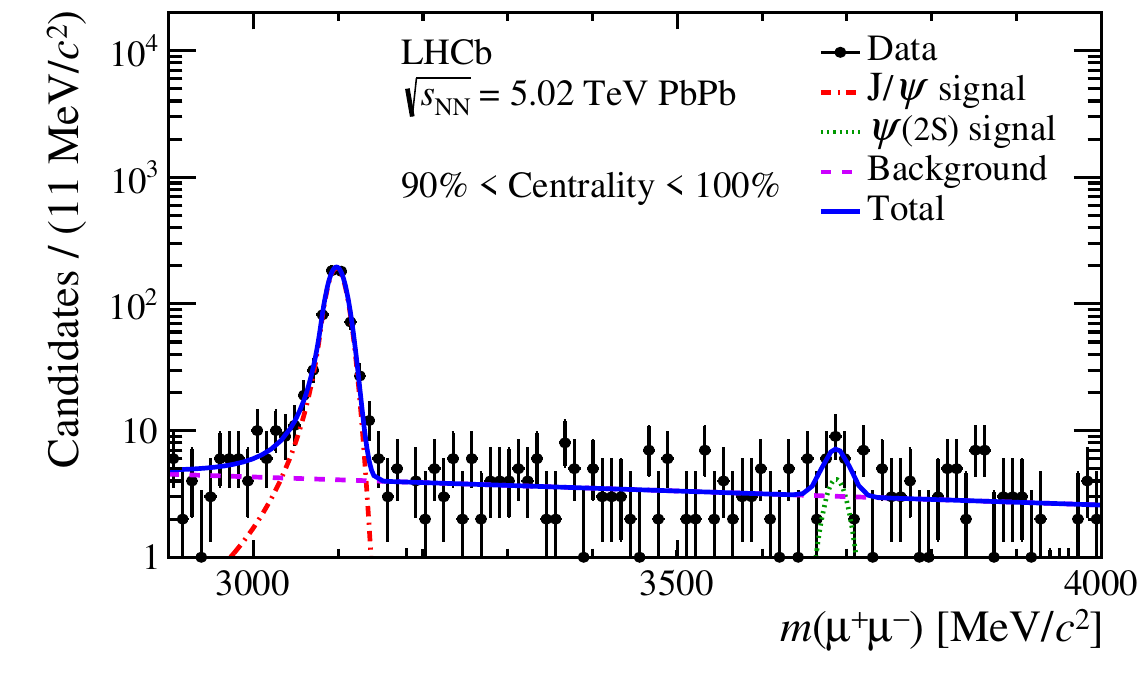}
    \vspace*{-0.5cm}
  \end{center}
  \caption{Dimuon invariant-mass distributions in each centrality interval after all selection requirements are applied, compared to the fit results.}
  \label{fig:MM}
\end{figure}

The charmonium yields, $N(\psitwos)$ and $N(\jpsi)$, are extracted by an unbinned maximum-likelihood fit to the dimuon invariant-mass distribution in each centrality interval where the \jpsi signal is modelled by a Crystal Ball function~\cite{Skwarnicki:1986xj} with tail parameters fixed from simulation. A Gaussian function is used to describe the \psitwos signal; the difference between the peak positions of the \jpsi and \psitwos signals is fixed to the difference of the known masses of the two states~\cite{PDG2024}. The background is described by an exponential function. 
\begin{table}[b]
    \centering
     \caption{\label{tab:mycen}Range for the number of participating nucleons $\langle N_{\rm{part}} \rangle$ and signal yields in each centrality interval, the uncertainties are statistical only.}
    \begin{tabular}{c|cccc}
         \hline
         Centrality (\%)&  90--100 &  80--90 &  70--80 &  60--70 \\
         \hline
         $\langle N_{\rm{part}} \rangle$& 2.4--5.5 & 5.5--13.0& 13.0--26.5&26.5--48.0\\
         $N(\jpsi)$& $596\pm28$ & $2099\pm52$ & $3320\pm74$ & $2221\pm77$\\ 
         $N(\psitwos)$& $13\pm5$ & \phantom{0}\phantom{0}$53\pm14$ & \phantom{0}\phantom{0}$68\pm26$ & \phantom{0}\phantom{0}$85\pm36$    \\
         \hline
    \end{tabular}
\end{table}
Figure~\ref{fig:MM} shows the dimuon invariant-mass distribution in each centrality interval compared to the fit results. The \jpsi and \psitwos yields extracted by the fit are shown in Table~\ref{tab:mycen}.
\section{Simulation and efficiencies}
\label{sec:eff}
The total signal selection efficiencies $\varepsilon_\mathrm{tot}$ in Eq.~\ref{eq:ratio} include the geometric acceptance of the detector, the efficiency due to event reconstruction, candidate selection,  particle identification and trigger requirements. Each contribution is determined using simulation and corrected for known differences between data and simulation.
The simulated decays are reconstructed and analysed using the same software tools as those used to process the data. In the simulation, \jpsi and \psitwos particles are generated using \pythia \cite{Sjostrand:2007gs} and embedded into minimum-bias PbPb collisions from the EPOS event generator \cite{Pierog_2015}. 
The decays of unstable particles are described by \evtgen \cite{Lange:2001uf}, in which final-state radiation is generated using PHOTOS \cite{Golonka_2006}. The interaction of the generated particles with the detector and its response are implemented using the GEANT4 toolkit \cite{Agostinelli:2002hh,Allison:2006ve}
adapted to LHCb~\cite{LHCb-PROC-2011-006}.
After reconstruction, the simulated events are weighted to improve agreement with data. The correction weights are obtained as a function of $\pt$ and $N_{\rm{c}}$ from the background-subtracted data and simulation sample of \jpsi candidates. The total efficiencies are estimated in each centrality interval, using the weighted simulation.

The acceptance efficiency, defined as the fraction of \jpsi \to \mup \mun decays with both muons entering the acceptance of the LHCb spectrometer, is determined using a simulation sample produced without any requirement on the
kinematics of the decay products. This efficiency is found to be independent of centrality. The combined reconstruction and selection efficiency, including the tracking,  is calculated as the fraction of generated events in the acceptance that pass the selection requirements other than those related to PID and the trigger.
The PID efficiency for the two muons is estimated from dedicated $pp$ calibration data samples~\cite{LHCb-PUB-2016-021}, and computed as a function of the muon kinematics. 
Specifically, the muon PID efficiency is determined in intervals of muon \pt and $N_{\rm{c}}$, showing a smooth dependence up to the edge of the multiplicity region at $N_{\rm{c}} = 5000$. In each muon \pt bin, it is then extrapolated via a quadratic fit function to $N_{\rm{c}}$ values up to 10000, to cover the multiplicity region of the measurement reported here using PbPb collisions.
Finally, the trigger efficiency is computed as the fraction of simulated signal particles that pass the trigger requirement.
\section{Systematic uncertainties}\label{sec:sys}
Several sources of systematic uncertainty associated with the determination of the signal and background yields and with the evaluation of the efficiencies are considered.

The systematic uncertainty arising from the signal parametrisation consists of two contributions. 
The first is associated with the choice of the fixed parameters of the \jpsi signal fit function. It is estimated by varying the parameters within their uncertainties obtained in simulation. 
The second contribution is estimated by repeating the signal yield extraction with an alternative parametrisation of the \jpsi signal using the convolution of a Crystal Ball and a Gaussian function where both widths are left free to vary.
Compared to the large statistical uncertainty affecting the \psitwos yield, the systematic uncertainty arising from the \psitwos fit model is negligibly small and ignored.
The uncertainty in background modelling is assessed by comparing the results from the baseline fit to those obtained using a second-order polynomial instead of the exponential function. As a cross-check, the baseline fit is repeated after constraining the slope of the exponential function using same-sign dimuon pairs, confirming the reliability of the uncertainty assigned from the background model variation.
The remaining nonprompt contamination in the candidate sample explained in Sec.~\ref{sec:datasel} is considered. The ratio measurement is repeated reducing the number of signal events by 8\% for the \psitwos and 5\% for the \jpsi; the difference with the baseline result is included as a systematic uncertainty on the ratio. 

The systematic uncertainty arising from the evaluation of the efficiencies consists of different contributions.
The first contribution is related to the weighting procedure applied to the simulation, and it is taken into consideration by randomly varying each weight within its uncertainty and estimating the total efficiencies again. Using 50 different sets of generated weights, the uncertainty considered is the relative variation of the obtained efficiencies. This contribution is convolved with the uncertainty associated with the procedure of applying the \jpsi weights to the \psitwos simulation. 
Another contribution pertains to the estimation of the PID efficiency in the high-multiplicity region that is not covered by the $pp$ calibration sample. The  extrapolation to that region is performed using a linear function instead of a quadratic function, and the variation of the efficiency is assigned as a systematic uncertainty.
The limited size of the simulation sample used to determine the efficiencies is also a source of systematic uncertainty, and is taken into account using the statistical uncertainties of the efficiencies determined from simulation.
Finally, a systematic uncertainty due to the trigger efficiency is assigned by using a data-driven method where the efficiency is determined from events triggered independently of the signal muons~\cite{LHCb-DP-2012-004}.
The total systematic uncertainty
is computed as the sum in quadrature of each contribution shown in Table~\ref{tab:sys}.
\begin{table}[tb]
    \centering
    \caption{Range of the relative systematic uncertainty contributions for the ratio measurement across the four centrality intervals.} 
    \label{tab:res_sys}
    \begin{tabular}{lc} 
    \hline
    Source & Uncertainty (\%) \\
    \hline
    Signal fit model & 0.5--2.2 \\
    Background fit model & 0.5--6.3 \\
    Nonprompt contamination & 2.8 \\
    Weighting procedure & 0.5 \\ 
    PID efficiency extrapolation & 0.1--2.8 \\
    Simulation sample size & 0.5--1.0 \\
    Trigger efficiency estimation & $<$ 0.1 \\
    \hline
    \end{tabular}
    \label{tab:sys}
\end{table}

\section{Results}
The ratio measurement integrated over the four centrality intervals is,
\begin{equation*}
\dfrac{\BF(\psitwos\to\mup\mun)}{\BF(J/\psi\to\mup\mun)}\cdot\dfrac{\sigma(\psitwos)}{\sigma(J/\psi)} = 0.022 \pm 0.006 \pm 0.001,
\end{equation*}
where the first and second uncertainties are statistical and systematic, respectively.

\begin{figure}[tb]
  \begin{center}
    \includegraphics[width=0.8\linewidth]{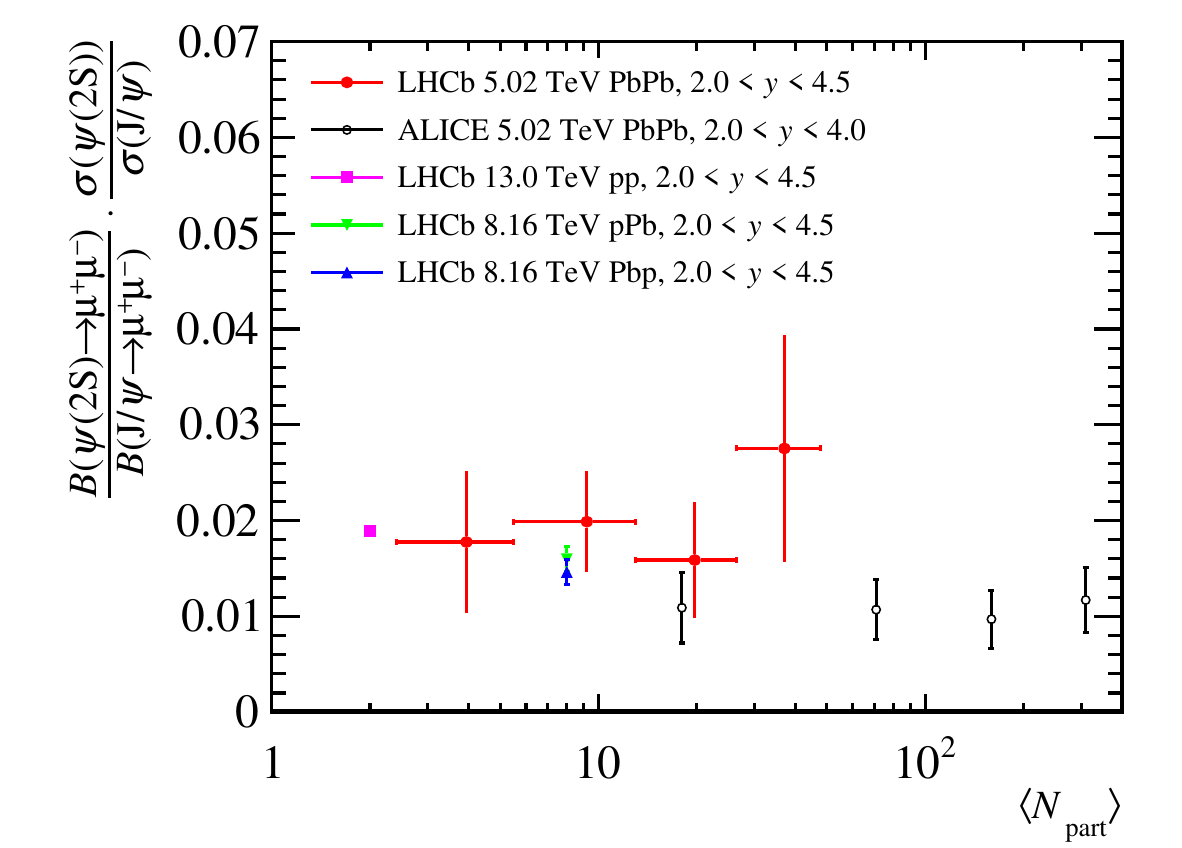}
    \vspace*{-0.5cm}
  \end{center}
  \caption{Cross-section times branching fraction ratio between $\ensuremath{\Ppsi{(2S)}}$ and $\ensuremath{{\PJ\mskip -3mu/\mskip -2mu\Ppsi}}$ states as a function of the number of participating nucleons $\langle N_{\rm{part}}\rangle$ in the \pt $>$ 0.3 \gevc region compared with ALICE PbPb~\cite{ALICE:2022jeh} and LHCb $pp$~\cite{LHCb-PAPER-2023-035} and $p$Pb~\cite{LHCb-PAPER-2023-024} results. The vertical error bars represent the sum in quadrature of the statistical and systematic uncertainties.}
  \label{fig:exp}
\end{figure}

\begin{figure}[tb]
  \begin{center}
    \includegraphics[width=0.8\linewidth]{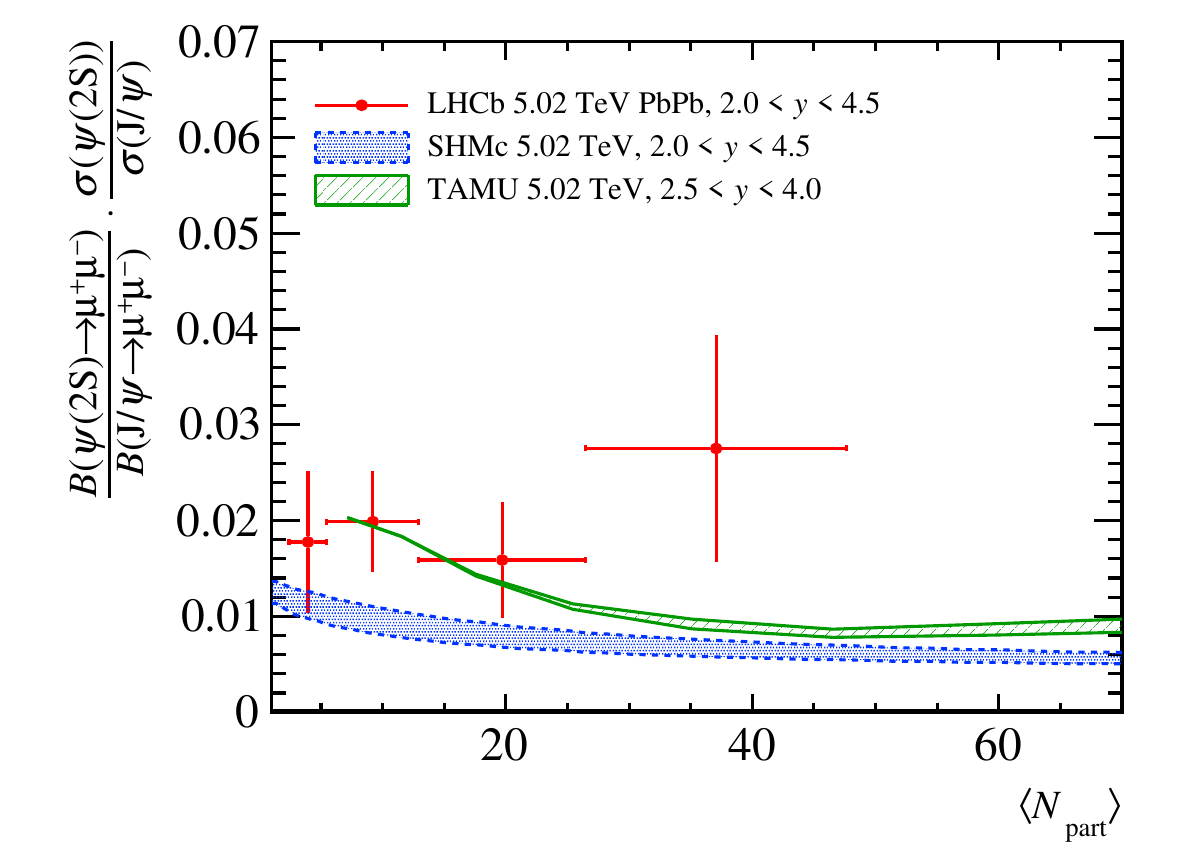}
    \vspace*{-0.5cm}
  \end{center}
  \caption{Cross-section times branching fraction ratio between $\ensuremath{\Ppsi{(2S)}}$ and $\ensuremath{{\PJ\mskip -3mu/\mskip -2mu\Ppsi}}$ states as a function of the number of participating nucleons $\langle N_{\rm{part}}\rangle$ in the \pt $>$ 0.3 \gevc region compared with the SHMc~\cite{Andronic:2019wva, Andronic:2017pug} and TAMU model~\cite{Du:2015wha,Wu:2024gil} predictions shown by the coloured bands representing the uncertainties. The vertical error bars represent the sum in quadrature of the statistical and systematic uncertainties.}
  \label{fig:th}
\end{figure}

The results in different centrality intervals are listed in Table~\ref{tab:res} and are presented as a function of the number of participating nucleons $\langle N_{\rm{part}} \rangle$ in Figs.~\ref{fig:exp} and \ref{fig:th}, where they are compared with some of the recently published measurements and with theory predictions, respectively.
In particular, Fig.~\ref{fig:exp} compares these results to the ALICE measurement~\cite{ALICE:2022jeh}, performed in PbPb collisions at the same centre-of-mass energy and at similar rapidities and transverse momenta but including both prompt and nonprompt contributions and covering a different centrality region. The two measurements are in agreement within uncertainties in the overlapping region, showing no dependency on $\langle N_{\rm{part}} \rangle$. The LHCb $pp$~\cite{LHCb-PAPER-2023-035} and $p$Pb~\cite{LHCb-PAPER-2023-024} measurements are reported in the same figure showing also agreement with the results reported here. The ratio measured in these studies is integrated over the kinematic range, and is associated with a number of participating nucleons of $\langle N_{\rm{part}}\rangle$ = 2 (8) for the $pp$ ($p$Pb) measurement, as estimated using the GMC model~\cite{Loizides:2014vua}.

Figure~\ref{fig:th} shows a comparison of these results with two recent theoretical calculations.
In the statistical hadronization model (SHMc)~\cite{Andronic:2019wva, Andronic:2017pug}, which includes regeneration at the phase boundary, it is assumed that, near the phase boundary between the QGP at high temperatures and the confined hadronic matter at lower temperatures, a fireball is formed during collisions, which is near thermal equilibrium.
The TAMU model~\cite{Du:2015wha,Wu:2024gil} is a transport model developed in a nonperturbative hydro-Langevin-RRM framework that includes collisional energy loss and heavy-quark diffusion in the medium and predicts recombination of charm quarks in the QGP phase.
When compared to the data reported in this article, the SHMc model, although showing no dependency of the ratio on the centrality, generally underestimates or is at the lower edges of its value. On the other hand, the TAMU model predictions are in better agreement with the experimental result, especially in the lower multiplicity intervals. 
A slight disagreement is observed in the highest multiplicity interval, where the data differ from the two theoretical predictions and the ALICE measurement by approximately 1.5 standard deviations.
It should be noted that while the TAMU model describes inclusive production, the SHMc model includes only the prompt component of charmonium production. In addition, the TAMU model has been computed on a slightly smaller rapidity range.

\begin{table}[b]
    \centering
     \caption{Cross-section times branching fraction ratio between \psitwos and \jpsi states in intervals of centrality and corresponding number of participating nucleons $\langle N_{\rm{part}}\rangle$. The first uncertainties are
    statistical and the second are systematic.}
    \begin{tabular}{ccc}
    \hline
         Centrality (\%)&  $\langle N_{\rm{part}} \rangle$ &  $\dfrac{\BF(\psitwos\to\mup\mun)}{\BF(J/\psi\to\mup\mun)}\cdot\dfrac{\sigma(\psitwos)}{\sigma(J/\psi)}$   \\
         \hline
         90--100 & 2.40--5.50 & $0.018 \pm 0.007 \pm0.001$  \\ 
         80--90  & 5.50--13.0 & $0.020 \pm 0.005 \pm0.001$  \\
         70--80  & 13.0--26.5 & $0.016 \pm 0.006 \pm 0.001$ \\
         60--70  & 26.5--48.0 & $0.027 \pm 0.011 \pm 0.001$ \\
    \hline     
    \end{tabular}
    \label{tab:res}
\end{table}
\clearpage
\section{Conclusions}
The \psitwos to \jpsi prompt production cross-section ratio in PbPb collisions at a  centre-of-mass energy of $5.02\tev$ corresponding to an integrated luminosity of around $230 \, \upmu\rm{b}^{-1}$ is measured as a function of event centrality. No significant dependence on the centrality is found.
The results are also compared to the latest theory predictions, with which they agree within 1.5 standard deviations or better, with a slight tension in the lowest centrality interval. The SHMc calculations tend to underestimate the data over the whole range; the TAMU model achieves a better description, albeit using a slightly different rapidity range. However, the current experimental precision precludes any firm conclusion. This study is a first step towards a deeper understanding of quarkonium production in semicentral PbPb collisions that can be achieved by additional measurements with the upgraded LHCb detector in Run 3~\cite{LHCb-DP-2022-002}, where, thanks to an improved tracking system, lower centrality values up to 30\% can be explored.

\section*{Acknowledgements}
%
%
\noindent We thank Anton Andronic, Ralf Rapp and Biao Gang for providing the theoretical predictions for our measurement.
We express our gratitude to our colleagues in the CERN
accelerator departments for the excellent performance of the LHC. We
thank the technical and administrative staff at the LHCb
institutes.
We acknowledge support from CERN and from the national agencies:
ARC (Australia);
CAPES, CNPq, FAPERJ and FINEP (Brazil); 
MOST and NSFC (China); 
CNRS/IN2P3 (France); 
BMBF, DFG and MPG (Germany); 
INFN (Italy); 
NWO (Netherlands); 
MNiSW and NCN (Poland); 
MCID/IFA (Romania); 
MICIU and AEI (Spain);
SNSF and SER (Switzerland); 
NASU (Ukraine); 
STFC (United Kingdom); 
DOE NP and NSF (USA).
We acknowledge the computing resources that are provided by ARDC (Australia), 
CBPF (Brazil),
CERN, 
IHEP and LZU (China),
IN2P3 (France), 
KIT and DESY (Germany), 
INFN (Italy), 
SURF (Netherlands),
Polish WLCG (Poland),
IFIN-HH (Romania), 
PIC (Spain), CSCS (Switzerland), 
and GridPP (United Kingdom).
We are indebted to the communities behind the multiple open-source
software packages on which we depend.
Individual groups or members have received support from
Key Research Program of Frontier Sciences of CAS, CAS PIFI, CAS CCEPP, 
Fundamental Research Funds for the Central Universities,  and Sci.\ \& Tech.\ Program of Guangzhou (China);
Minciencias (Colombia);
EPLANET, Marie Sk\l{}odowska-Curie Actions, ERC and NextGenerationEU (European Union);
A*MIDEX, ANR, IPhU and Labex P2IO, and R\'{e}gion Auvergne-Rh\^{o}ne-Alpes (France);
Alexander-von-Humboldt Foundation (Germany);
ICSC (Italy); 
Severo Ochoa and Mar\'ia de Maeztu Units of Excellence, GVA, XuntaGal, GENCAT, InTalent-Inditex and Prog.~Atracci\'on Talento CM (Spain);
SRC (Sweden);
the Leverhulme Trust, the Royal Society and UKRI (United Kingdom).





\addcontentsline{toc}{section}{References}
\bibliographystyle{LHCb}
\bibliography{main,standard,LHCb-PAPER,LHCb-CONF,LHCb-DP,LHCb-TDR}
  
\newpage
\centerline
{\large\bf LHCb collaboration}
\begin
{flushleft}
\small
R.~Aaij$^{38}$\lhcborcid{0000-0003-0533-1952},
A.S.W.~Abdelmotteleb$^{57}$\lhcborcid{0000-0001-7905-0542},
C.~Abellan~Beteta$^{51}$,
F.~Abudin{\'e}n$^{57}$\lhcborcid{0000-0002-6737-3528},
T.~Ackernley$^{61}$\lhcborcid{0000-0002-5951-3498},
A. A. ~Adefisoye$^{69}$\lhcborcid{0000-0003-2448-1550},
B.~Adeva$^{47}$\lhcborcid{0000-0001-9756-3712},
M.~Adinolfi$^{55}$\lhcborcid{0000-0002-1326-1264},
P.~Adlarson$^{82}$\lhcborcid{0000-0001-6280-3851},
C.~Agapopoulou$^{14}$\lhcborcid{0000-0002-2368-0147}
% ...

{\footnotesize \it

% ...
$^{14}$Universit{\'e} Paris-Saclay, CNRS/IN2P3, IJCLab, Orsay, France\\
% ...
$^{38}$Nikhef National Institute for Subatomic Physics, Amsterdam, Netherlands\\
% ...
$^{47}$Instituto Galego de F{\'\i}sica de Altas Enerx{\'\i}as (IGFAE), Universidade de Santiago de Compostela, Santiago de Compostela, Spain\\
% ...
$^{51}$Physik-Institut, Universit{\"a}t Z{\"u}rich, Z{\"u}rich, Switzerland\\
% ...
$^{55}$H.H. Wills Physics Laboratory, University of Bristol, Bristol, United Kingdom\\
% ...
$^{57}$Department of Physics, University of Warwick, Coventry, United Kingdom\\
% ...
$^{60}$School of Physics and Astronomy, University of Glasgow, Glasgow, United Kingdom\\
$^{61}$Oliver Lodge Laboratory, University of Liverpool, Liverpool, United Kingdom\\
% ...
$^{69}$Syracuse University, Syracuse, NY, United States\\
% ...
$^{82}$Department of Physics and Astronomy, Uppsala University, Uppsala, Sweden, associated to $^{60}$\\
% ...
}
\end{flushleft}

\end{document}